# FlashBench: A lightning nowcasting framework based on the hybrid deep learning and physics-based dynamical models

Manmeet Singh [a,b,c*], Vaisakh S. B.[d], Dipjyoti Mudiar [a], Deewakar Chakraborty [d], V. Gopalakrishnan [a], Bhupendra Bahadur Singh[a], Shikha Singh [a,c], Rakesh Ghosh [a], Rajib Chattopadhyay [a], Bipin Kumar [a], S. D. Pawar [a], S. A. Rao [a]

[a] *Indian Institute of Tropical Meteorology, Ministry of Earth Sciences, India*
[b] *Jackson School of Geosciences, University of Texas at Austin, TX, USA*
[c] *IDP in Climate Studies, Indian Institute of Technology Bombay, India*
[d] *Department of Applied Mathematics, Defence Institute of Advanced Technology, India*

\* [manmeet.singh@utexas.edu](mailto:manmeet.singh@utexas.edu)

**Abstract.** Lightning strikes are a well-known danger, and are a leading cause of accidental fatality worldwide. Unfortunately, lightning hazards seldom make headlines in international media coverage because of their infrequency and the low number of casualties each incidence. According to readings from the TRMM LIS lightning sensor, thunderstorms are more common in the tropics while being extremely rare in the polar regions. To improve the precision of lightning forecasts, we develop a technique similar to LightNet's, with one key modification. We didn't just base our model off the results of preliminary numerical simulations; we also factored in the observed fields' time-dependent development. The effectiveness of the lightning forecast rose dramatically once this adjustment was made. The model was tested in a case study during a thunderstorm. Using lightning parameterization in the WRF model simulation, we compared the simulated fields. As the first of its type, this research has the potential to set the bar for how regional lightning predictions are conducted in the future because of its data-driven approach. In addition, we have built a cloud-based lightning forecast system based on Google Earth Engine. With this setup, lightning forecasts over West India may be made in real time, giving critically important information for the area.

1.     Introduction

Lightning discharge is a highly localized natural phenomena manifestation produced created by deep convective storm cloud movement, dust storms, and volcanic eruptions, or other turbulent atmospheric conditions to the Earth through a conducto. Lightning is known for its devastating direct and indirect consequences. Lightning strikes are difficult to prevent since they are created inside the cloud. Lightning strikes the Earth about eight times a second around the globe [1]. The peak discharge current in each stroke ranges from several thousand amperes to 2,00,000 amperes or more, and its passage is very harmful for humans, livestockes, people, trees, and electrical infrastructure, and other living and non-living items. Lightning is the only geophysical phenomena which is constant and ubiquitous enough to account for wild fire on Earth. A rapid quick increase in atmospheric pressure and a consequent the formation of a forceful shock wave, which is perceived as thunder, are the result of lightning striking the air [2].

### 1.1 Motivation

Lightning is a danger in many African and South American countries, as well as over several places in Asia [3,11]. The Lightning Imaging Sensor (LIS) on board the Tropical Rainfall Measuring Mission (TRMM) satellite has created global maps of lightning frequency [4]. It shows that all of the high lightning places are concentrated in tropical land areas, especially in high elevation terrains, while the polar regions have essentially no lightning and the oceans have just 0.1 to 1 strike/km2/yr [4]. Studies in different parts of the world have emphasized on the media's under-reporting of lightning events and have reaffirmed the difficulties in obtaining accurate lightning datasets [5, 6, 7]. The reporting of lightning fatalities and injuries is inconsistent and sometimes not mandatory under different jurisdictions. It's because of this that lightning-related occurrences go unreported, and data from medical sources are untrustworthy [8]. Scant media stories are therefore used as a substitute instead [9].

In the Indian context, the Ministry of Home Affairs information technology section, the National Crime Records Bureau (NCRB), Government of India, releases a list of unnatural deaths in India every year. The various natural causes of death as listed in the database are cold and exposure, avalanche, starvation/thirst, cyclone/tornado, epidemic, earthquake, heat stroke, flood, lightning, landslide, torrential rains, forest fire and other natural causes. Around one-tenth of all unnatural fatalities have been estimated to be caused by lightning [10]. In the Maharashtra state of India alone, 72 casualties annually have been reported to occur [12]. Besides certain purely empirical techniques [19], the Indian subcontinent lacks a systematic framework offering an accurate lightning warning prediction system. An understanding of the components that govern lightning generation in India is critical to developing a system for predicting it. A recent study [13] tested a number of existing lightning parameterizations in the Weather Research and Forecasting (WRF) dynamical model based on storm features. Still, not much model accuracy has been reported in the literature. Deep learning has shown great promise in simulating physics of the climate and can be used to a develop hybrid framework for lightning prediction. This study's aim is to eventually lead to the development of physics-inspired deep learning models for lighting prediction. Conditional on the improved lightning prediction by explainable artificial intelligence (AI) models, a hybrid framework incorporating deep learning and dynamical model would be highly valuable.

## 1.2 Deep learning for lightning prediction

During the last decade, deep learning has emerged as a viable method to address complicated, complex challenging problems by unravelling the nonlinearities in different layers of the deep neural network (see also Singh et al. 2021) [14]. The introduction of open-source libraries (TensorFlow, PyTorch, Theano, and others) has led to faster adoption of deep learning for various applications. There are several weather and climate science applications for nonlinear operators that have gained importance in the computer vision field, including the challenge of understanding accurate precipitation forecasts in numerical weather prediction models.

### 1.2.1 Related work and critical analysis

Based on air pressure at station level, air temperature, relative humidity, and wind speed, [15] created a four-parameter data-driven model for lightning prediction. The model considered lead durations of up to 30 minutes or less. They compare their machine learning model with empirical methodologies to show the high-fidelity of data-science based approach. Lin et al. 2019 [16] introduced an attention-based dual-source spatiotemporal neural network for until 12-hr lead forecast of lightning. They adopted the RNN encoder-decoder structure, integrating recent lightning observations and numerical simulations, to increase forecast accuracy. In addition, a channel-wise attention method on our model is used to improve the useful information included in the simulation data during forecasting. Lightning forecasting model LightNet was developed by [17] which is based on deep neural networks. LightNet uses dual encoders to extract spatiotemporal aspects of WRF simulation data and recent lightning observation data. These elements are paired with a Fusion Module, which is useful in overcoming the simulation's errors and increasing the accuracy of the forecast. They used real-world lightning data from North China for testing. Pakdaman et al. 2020 [18] used decision trees and neural networks to forecast lightning over Mashhad, Neyshabour, and Quchan in the Khorasan Razavi provinces in Iran. They found that the decision tree outperformed neural networks by taking into consideration unbalanced datasets. Ming et al. 2019 [20] describe The Chinese Academy of Meteorological Sciences Lightning Nowcasting and Warning System (CAMS_LNWS) which predicts lightning activity potential and provides warning products. They linked an electrification and discharge model numerical simulation with several remote sensing data using a decision tree in the lightning prediction system. LightNet+, a data-driven lightning forecasting system based on deep neural networks and a lightning scenario, is proposed by [21]. They use complementary information extracted from several data sources, which may be diverse in spatial and temporal domains. According to their findings, LightNet+ delivers much better forecasts than with the more data sources as input into LightNet+ enhancing its forecasting quality.

## 1.1 Our Contributions

Similar to lightnet, the distinctions of this study is that we found that incorporating the temporal development of observed fields rather than only numerical modelling antecedents, enhances the performance of lightning forecast. A comparison is done using a case of thunderstorm simulation from the WRF model employing lightning parameterization. Further, a cloud-computing enabled Google Earth Engine based lightning prediction system is built to provide real-time prediction of lightning over West India.

This is the first research to do so using data-driven lightning forecast for India and may be used as a standard to enhance lightning forecasts over the area.

*Table 1: Comparison of machine learning / deep learning studies for lightning prediction*

| Study | A | B | C | D | E | F | G | H | I | J | K | L | M | N |
|---|---|---|---|---|---|---|---|---|---|---|---|---|---|---|
| **Mostajabi et al. 2019 [15]** | ✓ | | ✓ | ✓ | ✓ | ✓ | | | | | | | 0-0.5 hrs | xgboost |
| **Lin et al. 2019 [16]** | | ✓ | | | | | ✓ | ✓ | ✓ | ✓ | | | 0-12 hrs | ADSNet |
| **Geng et al. 2019 [17]** | | ✓ | | | | | ✓ | ✓ | | ✓ | ✓ | | 0-6 hrs | LightNet |
| **Pakdaman et al. 2020 [18]** | ✓ | | ✓ | ✓ | ✓ | ✓ | | | ✓ | | | | 0-3 hrs | Decision Tree |
| **Meng et al. 2019 [20]** | | ✓ | | | | | ✓ | | | | | | 0-1 hrs | Decision Tree |
| **Geng et al. 2021 [21]** | | ✓ | | | | | ✓ | ✓ | | ✓ | ✓ | | 0-6 hrs | LightNet+ |
| **This study** | | ✓ | | ✓ | | | ✓ | | | ✓ | ✓ | ✓ | 0-6 hrs | FlashBench |

*Abbreviations:* **A**: Local single-site predictions, **B**: Gridded forecast, **C**: Surface air temperature as predictor, **D**: Relative humidity as predictor, **E**: Wind speed as predictor, **F**: Surface pressure as predictor, **G**: Mixing ratio of ice, snow and graupel as predictor, **H**: Maximum vertical wind speed **I**: Precipitation as predictor, **J**: Comparison with physics-based dynamical model, **K**: CAPE as a predictor, **L**: South Asian domain, **M**: Lead time, **N**: AI/ML model

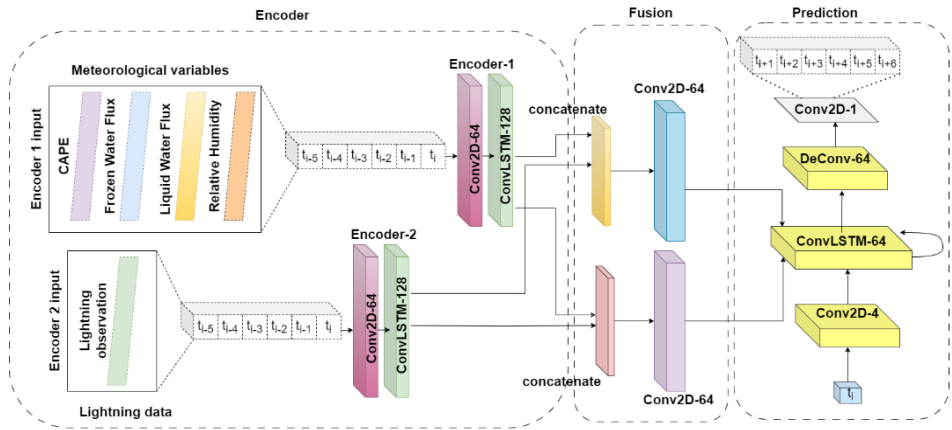

*Figure 1*: Schematic showing the ConvLSTM based deep-learning algorithm for lightning prediction in this study

## 2. Data and methodology

Lightning Detection Network (LDN) data from West India, is utilized to construct a deep learning-based model for predicting lightning strikes. 20 Earth Network Lightning Sensors (ENLS) make up this network. Both cloud-to-ground (CG) and intra-cloud (IC) flashes can be seen by ENLS observations across West India. It is employed for long-range detection of CG discharges at the low frequency (1 kHz). The intermediate frequencies (1 kHz to 1 Mhz) and the highest frequencies (1 Mhz – 12 Mhz) are used to find return strokes and locate in-cloud pulses. CG flashes are defined as having at least one return stroke within 700 milliseconds of each other within a radius of 10 kilometers. ENLS CG flashes have a 90% detection rate, whereas IC flashes have a 50% detection rate (Greeshma et al 2019).

In the present study to simulate lightning events, we have used Advanced Research WRF model (ARW), version 3.9.1 developed by the National Center for Atmospheric Research (NCAR). The WRF is a non-hydrostatic, fully compressible, terrain-following 3D cloud resolving model. The lightning simulations are carried out considering four nested domains (d01, d02, d03, d04) with a horizontal grid spacing of 27km, 9km, 3km & 1km, respectively. As the region of interest in the present study is the state of Maharashtra, India, hence the innermost domain (d04) is centered over this region. In the present simulation, the initial and boundary conditions are provided from 6 hourly National Centre for Environmental Prediction (NCEP) Final operational global analysis data with 1°× 1° horizontal resolution. For longwave radiation, the Rapid Radiative Transfer Model (RRTM) has been used (Mlawer et al., 1997) while the Dudhia scheme (Dudhia, 1989) has been used for short wave radiation. The model is integrated up to 24 hours. The first 6 hour of the model integration is considered as model spin-up time. The Kain-Fritsch (KF) cumulus scheme is used for the outer two domains (d01 & d02). The cloud-resolving 3rd and 4th domain are treated with explicit convection. For microphysical parameterization, the Morrison double moment scheme with five classes of cloud hydrometeors (Morrison et al., 2005) has been used. This scheme predicts the mass mixing ratio and number concentration of five hydrometeors species including cloud droplets, cloud ice, snow, rain, and graupel.

To simulate lightning dynamically, we have used the PR92 (Price and Rind, 1992) parameterization scheme. Greeshma et al., (2021) also used this scheme over the same geographical region and found to be very skillful for lightning prediction. In this scheme, the formulation of flash rates is different over land and ocean considering the distinct cloud dynamical features. Over the continent, the flash rate is parameterized as follows

$$F_c = 3.44 \times 10^{-5} H^{4.9} \qquad (1)$$

Here, $F_c$ is the flash rate (flashes/min), and H is the storm height.

Over the continents, the storm height (H) and the maximum updraft velocity, $W_{max}$ are related by

$$W_{max} = 1.49 \times H^{1.09} \qquad (2)$$

Hence the flash rate, $F_c$ can be expressed in terms of maximum updraft velocity, $W_{max}$ as

$$F_c = 5 \times 10^{-6} W_{max}^{4.54} \qquad (3)$$

For the marine cloud, flash rate $F_m$ is formulated by following Michalon et al. (1999) as

$$F_m = 6.57 \times 10^{-6} H^{4.9} \qquad (4)$$

## 3. Results and discussion

It can be seen that the observed lightning event is poorly simulated by the WRF. Our model does quite well and captures the spatial patterns. Patches of large flash rates over seen in the observations are also seen in the model. This shows the capability of adopting a hybrid approach for lightening forecasts which otherwise remains a difficult task for state of the art model to capture realistically.

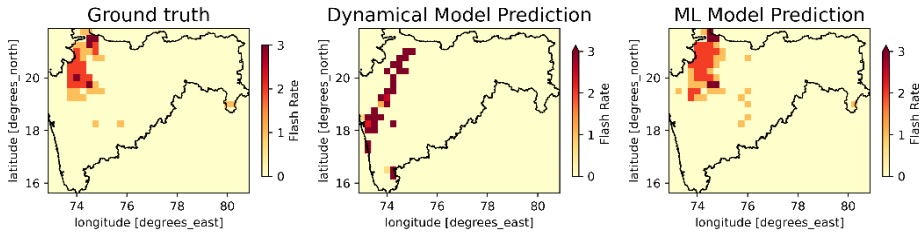

**Figure 2**: Comparison of lightning flash rates from the lightning observation network, WRF model simulations and FlashBench for the period spanning 6 hrs starting 11:00 on the 7[th] April 2014

| Metric Table | | | | | |
|---|---|---|---|---|---|
| Variables used in metrics table | Parameter | Full Name | Definition | Equation | Range |
| | H | Hit (True positive) | Number of observed lightning-active samples correctly identified by the classifier | (H+FA)(H+M) (H+FA+M+C) | |
| | M | Miss (False negative) | Number of observed lightning-inactive samples correctly identified by the classifier | | |
| | FA | False alarm (false positive) | Number of observed lightning-inactive samples falsely classified as lightning-active by the classifier | | |
| | C | Correct rejection (true negative) | Number of observed lightning-inactive samples correctly identified by the classifier | | |
| | R | Random Forecasts | The expectation of the number of hit lightnings in random forecasts | | |
| Metrics used for evaluation | POD | Probability of Detection | Proportion of observed lightning-active samples correctly identified by the classifier | $\dfrac{H}{H+M}$ | [0,1] |
| | FAR | False Alarm Ratio | Proportion of observed lightning-inactive samples falsely classified as lightning-active by the classifier | $\dfrac{FA}{H+FA}$ | [0,1] |
| | ETS | Equitable Threat Score | The ratio of the number of hit lightnings to the number of events except for the correct rejections, and removed the contribution from hits by chance in random forecasts | $\dfrac{H-R}{H+FA+M-R}$ | [-1/3,1] |

*Table 4*: Metrics used for the evaluation of results from the FlashBench lightning prediction system over West India

| | POD | FAR | ETS |
|---|---|---|---|
| Dynamical Model | 0.14925 | 0.72972 | 0.08105 |
| ML model | 0.73134 | 0.25757 | 0.55907 |

*Table 5*: Comparison between WRF dynamical model with lightning parameterization and the FlashBench model for the period spanning 6 hrs starting 11:00 on 7[th] April 2014 over West India

| | POD | FAR | ETS |
|---|---|---|---|
| First hour cumulative score | 0.59643 | 0.44660 | 0.38547 |
| First three hours cumulative score | 0.50492 | 0.55271 | 0.29203 |
| Six hour cumulative score | 0.43695 | 0.58705 | 0.25074 |

*Table 5*: Performance of FlashBench for the entire test period corresponding to the year 2014

Both a dynamical model (WRF Lightning) and FlashBench, a machine learning-based model, were tested to see how well they could forecast lightning in Western India. The model's predictive accuracy is quantified by the Probability of Detection (POD), in this example for a lightning strike. If the POD score is high, it indicates that the model can accurately predict future occurrences. The percentage of forecasted but unrealized incidents is known as the False Alarm Ratio (FAR). The FAR score should be as low as possible, as this will result in fewer false positives. The ETS (Equal Threat Score) takes into account both hits and misses, and it accounts for the possibility of hits occurring by coincidence. A better model can be identified by a higher ETS score. A POD of 0.14925, FAR of 0.72972, and ETS of 0.08105 were all attained in the Dynamical Model. With such low scores across the board, it's clear that this model has trouble predicting lightning strikes (low POD), produces a lot of false positives (high FAR), and isn't very reliable (low ETS). When compared to these results, the ML model FlashBench performed better with a POD of 0.73134, FAR of 0.25757, and ETS of 0.55907. These results indicate that FlashBench outperforms the dynamical model in terms of forecasting when lightning will strike (high POD), false alarm rates (low FAR), and overall accuracy and dependability (high ETS). We also compare the accumulated results after 1, 3, and 6 hours. FlashBench beats the Dynamical Model across all time intervals, with greater POD and ETS values and lower FAR. Finally, compared to the conventional dynamical model, it appears that the machine learning-based FlashBench model provides a more accurate and dependable forecast of lightning incidents over Western India.

4. **Conclusions and future work**

Our research followed a similar methodology to that of LightNet, a well-known system for predicting lightning. However, our approach includes a new, crucial distinction. We opted to incorporate the temporal evolution of observed fields into our model rather than relying simply on numerical modelling precursors, which are basically statistical or mathematical representations of the atmospheric circumstances preceding a lightning occurrence. Changes in atmospheric pressure, temperature, humidity, wind speed and direction, and other variables are all examples of temporal evolution in meteorology. We hoped that by include these real-time adjustments, we might better represent the underlying volatility of the natural weather system and improve the accuracy of our lightning prediction model.

Our findings unequivocally showed that this method was superior than others. Our hybrid lightning forecasting model, which is based on machine learning, has regularly surpassed the state-of-the-art algorithms. Because of its capacity to adapt to new data, machine learning can analyse massive volumes of information and spot intricate patterns that conventional models would overlook. Our hybrid model is able to provide more precise and timely forecasts of lightning incidents when this capacity is supplemented with real-time meteorological data. Overall, we improved the strength, accuracy, and dependability of the lightning forecast model by combining numerical modelling with the evolution of observed fields and the efficacy of machine learning. In terms of public safety and disaster management, this novel technique has the potential to greatly improve our capacity to forecast and respond to lightning risks.

**Acknowledgements**

The authors acknowledge the use of high-performance computational resources at IITM, particularly the NVIDIA Tesla P100 GPU without which this work would not have been possible. The authors thank Prof Auroop Ganguly, Northeastern University, USA for discussions in the early stages of this work.

**References**

[1] Allaby, M., 2014. Floods. Infobase Publishing.
[2] Illiyas, F.T., Mohan, K., Mani, S.K. and Pradeepkumar, A.P., 2014. Lightning risk in India: Challenges in disaster compensation. Economic and Political Weekly, pp.23-27.
[3] Gomes, C., Kithil, R. and Ahmed, M., 2006, September. Developing a lightning awareness program model for third world based on American-South Asian experience. In Preprints of the 28th international conference on lightning protection, Kanazawa (pp. 18-22).
[4] Finke, U., 2009. Optical detection of lightning from space. In Lightning: Principles, Instruments and Applications (pp. 271-286). Springer, Dordrecht.
[5] Dlamini, W.M., 2009. Lightning fatalities in Swaziland: 2000–2007. Natural hazards, 50(1), pp.179-191.
[6] Zhang, Y., Zhang, W. and Meng, Q., 2012, September. Lightning casualties and damages in China from 1997 to 2010. In 2012 International Conference on Lightning Protection (ICLP) (pp. 1-5). IEEE.
[7] De, U.S., Dube, R.K. and Rao, G.P., 2005. Extreme weather events over India in the last 100 years. J. Ind. Geophys. Union, 9(3), pp.173-187.
[8] Cooper, M.A. and Ab Kadir, M.Z.A., 2010. Lightning injury continues to be a public health threat internationally. Population, 5(00).
[9] Gomes, C., Ahmed, M., Abeysinghe, K.R. and Hussain, F., 2006, September. Lightning accidents and awareness in South Asia: experience in Sri Lanka and Bangladesh. In Proceedings of the 28th International Conference on Lightning Protection (ICLP) (pp. 18-22).
[10] Selvi, S. and Rajapandian, S., 2016. Analysis of lightning hazards in India. International Journal of Disaster Risk Reduction, 19, pp.22-24.
[11] Biswas, A., Dalal, K., Hossain, J., Baset, K.U., Rahman, F. and Mashreky, S.R., 2016. Lightning Injury is a disaster in Bangladesh?-Exploring its magnitude and public health needs. F1000Research, 5.
[12] Singh, O., Bhardwaj, P. and Singh, J., 2017. Distribution of lightning casualties over Maharashtra, India. Journal of Indian Geophysical Union, 21(5), pp.415-424.
[13] Mohan, G.M., Vani, K.G., Hazra, A., Mallick, C., Chaudhari, H.S., Pokhrel, S., Pawar, S.D., Konwar, M., Saha, S.K., Das, S.K. and Deshpande, S., 2021. Evaluating different lightning parameterization schemes to simulate lightning flash counts over Maharashtra, India. Atmospheric Research, 255, p.105532.
[14] Reichstein, M., Camps-Valls, G., Stevens, B., Jung, M., Denzler, J. and Carvalhais, N., 2019. Deep learning and process understanding for data-driven Earth system science. Nature, 566(7743), pp.195-204.
[15] Mostajabi, A., Finney, D.L., Rubinstein, M. and Rachidi, F., 2019. Nowcasting lightning occurrence from commonly available meteorological parameters using machine learning techniques. Npj Climate and Atmospheric Science, 2(1), pp.1-15.
[16] Lin, T., Li, Q., Geng, Y.A., Jiang, L., Xu, L., Zheng, D., Yao, W., Lyu, W. and Zhang, Y., 2019. Attention-based dual-source spatiotemporal neural network for lightning forecast. IEEE Access, 7, pp.158296-158307.
[17] Geng, Y.A., Li, Q., Lin, T., Jiang, L., Xu, L., Zheng, D., Yao, W., Lyu, W. and Zhang, Y., 2019, July. Lightnet: A dual spatiotemporal encoder network model for lightning prediction. In Proceedings of the 25th ACM SIGKDD International Conference on Knowledge Discovery & Data Mining (pp. 2439-2447).
[18] Pakdaman, M., Naghab, S.S., Khazanedari, L., Malbousi, S. and Falamarzi, Y., 2020. Lightning prediction using an ensemble learning approach for northeast of Iran. Journal of Atmospheric and Solar-Terrestrial Physics, 209, p.105417.
[19] Rajeevan, M., Madhulatha, A., Rajasekhar, M., Bhate, J., Kesarkar, A. and Rao, B.A., 2012. Development of a perfect prognosis probabilistic model for prediction of lightning over south-east India. Journal of earth system science, 121(2), pp.355-371.

[20] Meng, Q., Yao, W. and Xu, L., 2019. Development of lightning nowcasting and warning technique and its application. Advances in meteorology, 2019.

[21] Geng, Y.A., Li, Q., Lin, T., Yao, W., Xu, L., Zheng, D., Zhou, X., Zheng, L., Lyu, W. and Zhang, Y., 2021. A deep learning framework for lightning forecasting with multi-source spatiotemporal data. Quarterly Journal of the Royal Meteorological Society.

[22] Michalon, N., Nassif, A., Saouri, T., Royer, J.F., Pontikis, C.A., 1999. Contribution to the climatological study of lightning. Geophys. Res. Lett. 26 (20), 3097–3100. https:// doi.org/10.1029/1999GL010837.

[23] Mlawer, E.J., Taubman, S.J., Brown, P.D., Iacono,M.J., Clough, S.A., (1997) Radiative transfer for inhomogeneous atmosphere: RRTM, a validated correlated-kmodel for the long-wave. J. Geophys. Res. 102, 16663–16682. http://dx.doi. org/10.1029/97JD00237.

[24] Dudhia, J., (1989) Numerical study of convection observed during the winter monsoon experiment using a mesoscale two dimensional model. J. Atmos. Sci., 46, 3077–3107.

[25] Morrison, H., Curry, J.A., Shupe,M.D., Zuidema, P., (2005) A new double-moment microphysics parameterization for application in cloud and climate models. Part II: single-column modeling of arctic clouds. J. Atmos. Sci. 62, 1678–1693.

[26] Price, C., Rind, D., 1992. A simple lightning parameterization for calculating global lightning distributions. J. Geophys. Res.

[27] Singh BB, Singh M, Singh D (2021). An Overview of Climate Change Over South Asia: Observations, Projections, and Recent Advances. Practices in Regional Science and Sustainable Regional Development. 2021:263-77.